\documentclass[toc]{PoS}
\usepackage{graphicx}
\usepackage{amssymb}
\usepackage{fontenc}
\usepackage{times}
\usepackage{mathptmx}

\newcommand{\be}{\begin{equation}}
\newcommand{\ee}{\end{equation}}
\newcommand{\bea}{\begin{eqnarray}}
\newcommand{\eea}{\end{eqnarray}}

\def\slash#1{\setbox0=\hbox{$#1$}#1\hskip-\wd0\dimen0=5pt\advance
       \dimen0 by-\ht0\advance\dimen0 by\dp0\lower0.5\dimen0\hbox
         to\wd0{\hss\sl/\/\hss}}

\title{QCD critical point: a historical perspective}

\ShortTitle{QCD critical point:}

\author{{Roberto Casalbuoni}\\
        Dipartimento di Fisica dell'Universita' di Firenze and Sezione INFN\\
        Via G. Sansone 1, 50019 Sesto Fiorentino (FI), Italy\\
        E-mail: \email{casalbuoni@fi.infn.it}}

\abstract{We review the history of the critical point in the QCD
phase diagram, baryon density vs. temperature. In particular we
discuss the different theoretical approaches to this problem.}

\FullConference{The 3rd edition of the International Workshop ---
The Critical Point and Onset of Deconfinement --- \\
         July 3-7 2006\\
         Galileo Galilei Institute, Florence, Italy}

\begin{document}

\section{Introduction}

In order to understand the properties of ordinary matter (baryons
and mesons) it is necessary to understand the properties of the
ground state of QCD. The best way to test the physical properties of
a system is to vary its defining conditions    in order to test its
reactions. To do that in QCD we have to consider a system different
from the vacuum but sufficiently simple in order to be able to study
it. Some possibilities are to study extremely dense matter  or
matter at  very high temperature as at the beginning  of our
universe. A bonus in studying the ground state of these particular
systems is that, due to the fundamental property of asymptotic
freedom, QCD simplifies a lot. In the case of very high temperature
one expects  QCD to behave as a free theory describing a non
interacting gas of quarks and gluons. A similar conclusion would be
true also at high density except that having to do with fermions we
have to keep into account the exclusion principle. As a consequence
a very degenerate Fermi sphere is formed and, if an arbitrary
attractive interaction is present, we expect that the phenomenon of
color superconductivity takes place \cite{barrois,cs}. This is what
we expect in QCD, since at very high density the theory can be
described in terms of a single gluon exchange and this provides an
attraction in the antisymmetric diquark state. From this we get
informations about two asymptotic regions of the phase diagram of
QCD in the variables $(\mu, T)$, where $\mu$ is the baryon chemical
potential. The two regions are respectively $(\mu\approx 0, T\to
\infty)$ and $(\mu\to\infty, T\approx 0)$. However we would like
also to know what happens in  the intermediate region and in
particular we would like to determine  the order of the phase
transitions from the hadronic phase to the quark-gluon plasma and to
the color superconducting phase. In this context  the possibility
arises that  going from the hadronic  to the gluon-quark plasma
phase, there is a cross-over for small chemical potentials and a
first order for higher values of $\mu$. In this case, the end point
of the first-order line is  called the "critical point" of QCD. The
study of this intermediate region is quite complicated since
perturbation theory cannot be applied to QCD and furthermore at
finite chemical potential the usual lattice approach fails. In this
talk I will discuss the historical path through which the idea of a
critical point came about and some of the attempts of locating it.
Therefore I will not discuss  the physics associated to the critical
point and how this point can be detected experimentally. Both these
topics are treated by other speakers in this conference and   many
nice reviews about the subject exist in the literature, see for
example
\cite{Rajagopal:1995bc,Rajagopal:2000wf,Stephanov:2004wx,Satz:2005md,Schafer:2005ff}.

\section{Order parameters}

The quark gluon plasma phase can be thought of as a deconfined phase
and therefore one would like to define an order parameter to
distinguish between this and the confined phase (the hadronic one).
Such an order parameter can be easily defined for a theory without
quarks (or, equivalently for $m_q\to\infty$). This is the Polyakov
loop \cite{polyakov} defined as \be L(\vec x)={\rm tr}\,\Omega(\vec
x),~~~\Omega(\vec x)= P \exp\left(i\int_0^\beta dt A_0(\vec x,
t)\right)\ee with $\beta= 1/{kT}$ and $A_0$ the time component of
the gluon field. It turns out that the expectation value of the loop
is asymptotically given by \be \langle
L\rangle\approx\lim_{r\to\infty} e^{-\beta V(r)}\ee with $V(r)$ the
potential between a static quark-antiquark pair at a distance $r$.
Therefore the confined and the deconfined phase are distinguished by
the value of $\langle L\rangle$ \bea {\rm Confined~ phase:}&&\langle
L\rangle =0\cr {\rm Deconfined~ phase:}&&\langle L\rangle \not=0\eea
From a symmetry point of view, $L$ characterizes the breaking of the
center of the color group, $Z(N_c)$, in the case of $N_c$ colors.
From asymptotic freedom we expect that at some critical temperature
$T_c$ \be  \langle L\rangle=0,~~T<T_c,~~~\langle
L\rangle\not=0,~~T>T_c\ee For finite quark masses we expect $V(r)$
to remain finite for $r\to\infty$. In fact the string of the color
flux between the two color charges is expected to break when the
potential energy equals the mass of the lowest hadronic state,
$M_h$. Therefore $\langle L\rangle$ does not vanish in the hadronic
phase but rather goes exponentially to zero for $M_h\to\infty$ \be
\langle L\rangle\approx e^{-\beta M_h}\ee
\begin{figure}[h]\begin{center}
\includegraphics[width=.6\textwidth]{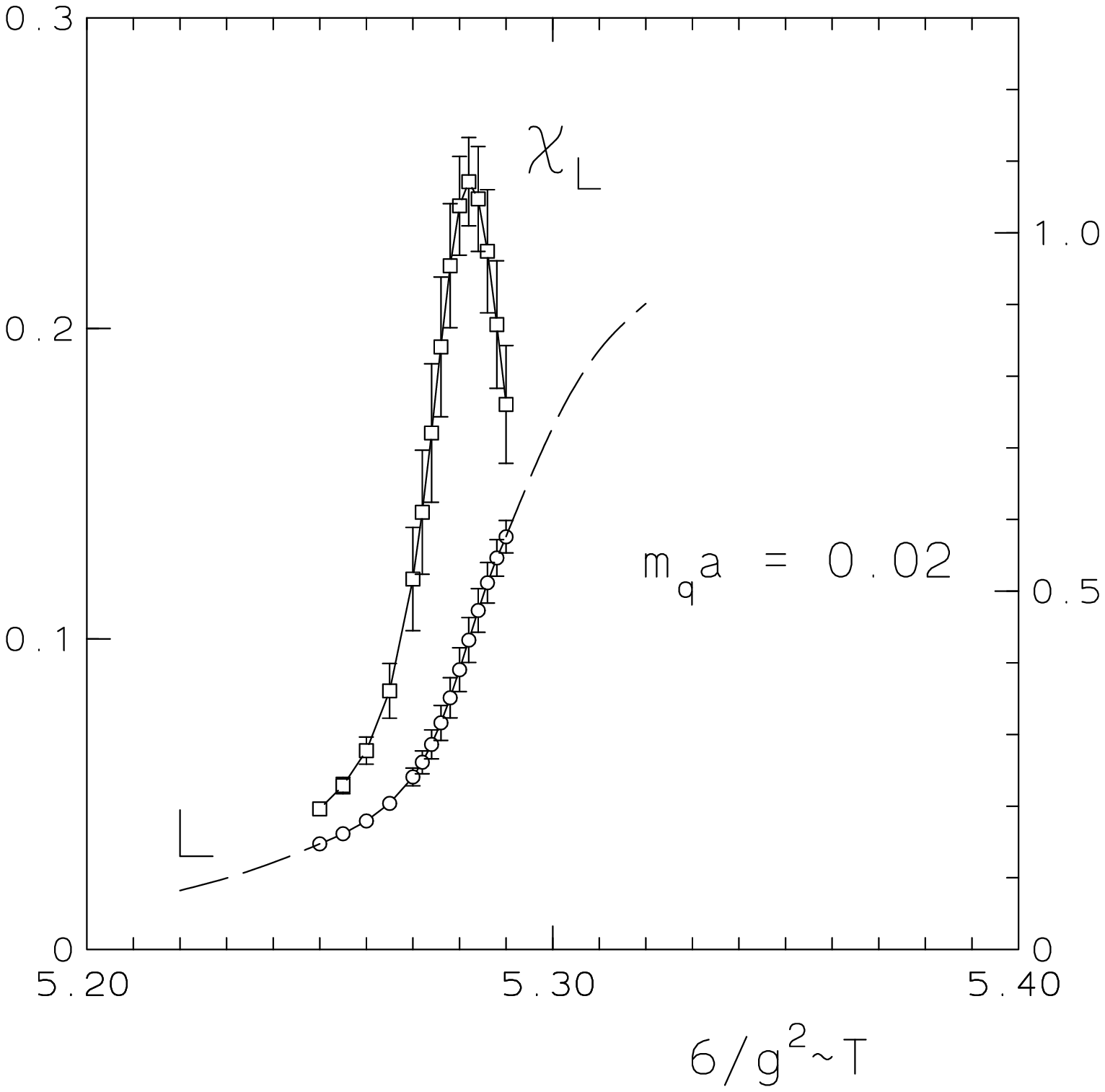}
\includegraphics[width=.6\textwidth]{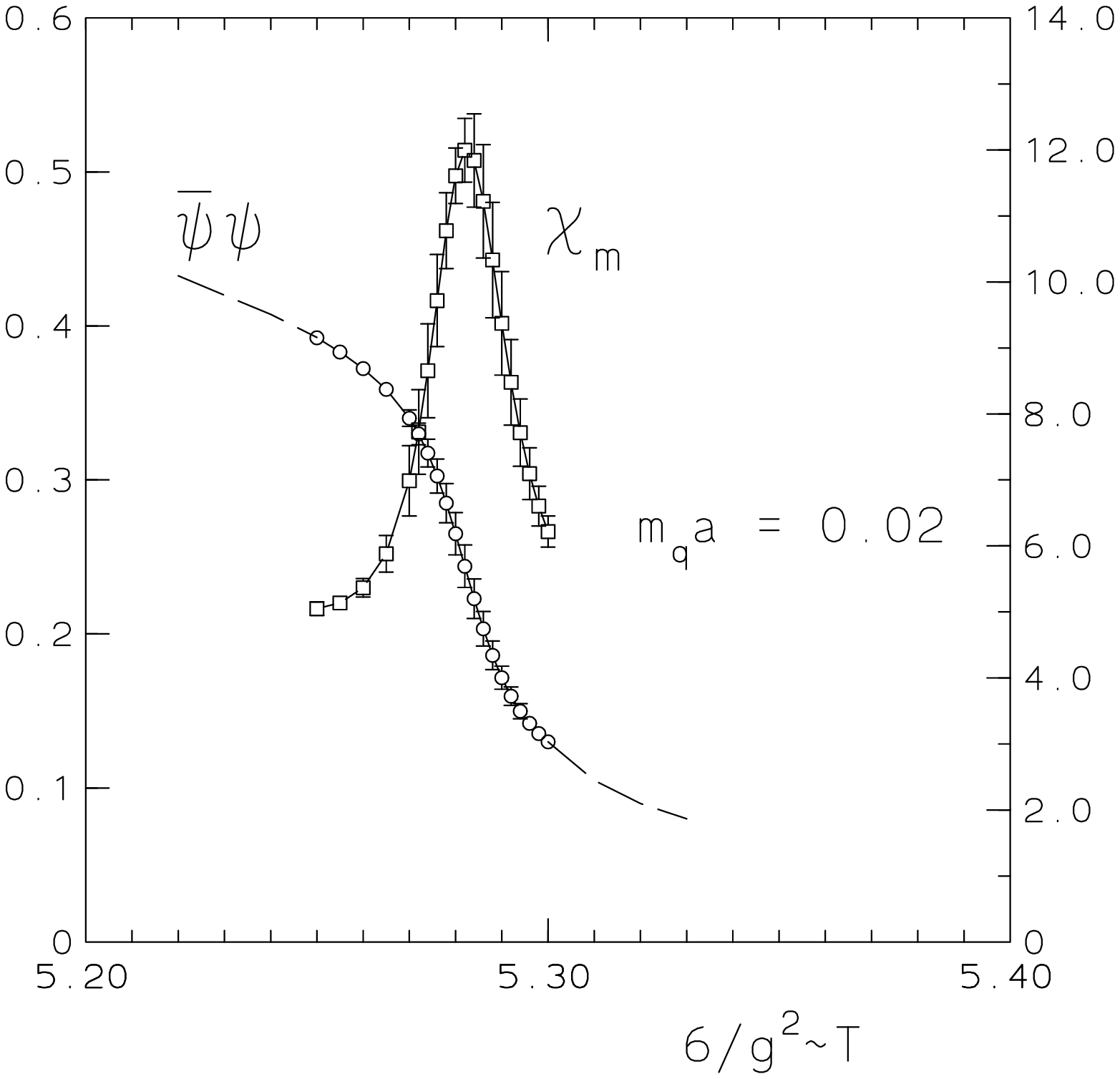}\end{center}
\caption{The Polyakov looop and the chiral condensate
susceptibilities in the case of two flavor QCD. This calculation has
been made with a quark mass about four times bigger than the one
needed for obtaining the physical pion mass \cite{Karsch:1994hm}.}
\label{fig1}
\end{figure}
When quarks are present one can define another order parameter, the
chiral condensate, $ \langle\bar\psi\psi\rangle$, characterizing the
breaking of the flavor symmetry (for instance, for three massless
flavors the chiral symmetry would be $SU(3)_L\otimes SU(3)_R\otimes
U(1)_V$). In this case we expect \be \langle\bar\psi\psi\rangle =
0~~{\rm for}~~ T\to \infty,~~~ \langle\bar\psi\psi\rangle\not =
0~~{\rm for}~~ T\to 0\ee Of course, since $m_q\not=0$, this order
parameter never vanishes but it will have a sharp variation, or a
crossover, close to the transition. The susceptibilities of these
two order parameters \be\chi_L\approx \langle L^2\rangle-\langle
L\rangle^2,~~~\chi_m \approx\frac{\partial
\langle\bar\psi\psi\rangle}{\partial m_q}\ee have been evaluated on
the lattice \cite{Karsch:1994hm} in the case of two flavors. The
results are shown in Fig. \ref{fig1}. The figure shows very clearly
that the deconfinement and the chiral transition coincide at zero
baryon density.

Our final conclusion is that the phase structure is characterized by
\bea &T<T_c~~{\rm confined~phase:}~~~ \langle L\rangle\approx
0,~~\langle\bar\psi\psi\rangle\not=0&\cr &T>T_c~~{\rm
deconfined~phase:}~~~ \langle L\rangle\not=
0,~~\langle\bar\psi\psi\rangle\approx 0&\eea Given this result, in
the following we will concentrate on the chiral transition which is
easier to deal with.

\section{First attempts to evaluate the phase diagram of QCD}

One of the first attempts to evaluate the $(\mu, T)$ phase diagram
of QCD was done in ref. \cite{Bailin:1984ak}. The authors evaluated
the gap equation for the chiral condensate in the approximation of
one gluon-exchange. However the paper did not contain a discussion
about the nature of the chiral transition. Other attempts
\cite{Kocic:1985uq,Galina:1987bj} were done using the Coulomb gauge
and neglecting the retardation effects in the gluon propagator. This
approach is simple since it is very close to a non-relativistic
treatment and it allows to vary the static potential according to
the assumptions for the gluon exchange. For instance, the static
potential has been chosen as a $\delta$-function, Coulomb type or
confining. In all these cases these authors have found a second
order transition in the plane $(\mu,T)$.

A completely different approach was developed in refs.
\cite{Damgaard:1985bn, Ilgenfritz:1984ff}. The authors derived
effective lagrangians for different gauge groups and a single
flavor. In particular they found that for $N_c=2$ the line of
transition is second order, whereas for $N_c=3$ it is first order.

The paper in ref. \cite{Asakawa:1989bq} started the analysis of the
problem by using a Nambu-Jona Lasinio (NJL) model. The case studied
was $N_f=2$ and $N_c=3$. The idea is to simulate the gluon
interaction through an effective four-fermi coupling. Although there
is no reason to expect quantitative results close to  real QCD, one
hopes that universal  effects can be recovered. In ref.
\cite{Asakawa:1989bq} the interaction lagrangian used is \bea &{\cal
L}_I={\cal L}_{sym}+{\cal L}_{det}&\cr & {\cal L}_{sym}=\frac 1 2
g_1\left[(\bar\psi\psi)^2+(\bar\psi i\gamma_5\vec\tau\psi)^2
+(\bar\psi i\gamma_5\psi)^2+(\bar\psi\vec \tau\psi)^2\right]&\cr
&{\cal L}_{det}=\frac 1 2 g_2\left[(\bar\psi\psi)^2+(\bar\psi
i\gamma_5\vec\tau\psi)^2 -(\bar\psi i\gamma_5\psi)^2-(\bar\psi\vec
\tau\psi)^2\right]&\eea where ${\cal L}_{det}$ is the t'Hooft
determinant breaking the axial symmetry $U(1)_A$, written for
$N_f=2$. In general \be {\cal L}_{det}=\frac 1 2 g_2 \left[{\rm
det}\{\bar\psi(1+\gamma_5)\psi\}+h.c.\right]\ee For simplicity the
authors did the choice $g_1=g_2=g$. Therefore the model depends on 3
parameters, the mass of the quarks $m$, the coupling $g$ and the
cutoff $\Lambda$ defining the model. These parameters can be
determined at zero temperature and density by using the physical
values of $m_\pi$, $f_\pi$ and a reasonable value for the
condensate.
\begin{figure}[h]\begin{center}
\includegraphics[width=.7\textwidth]{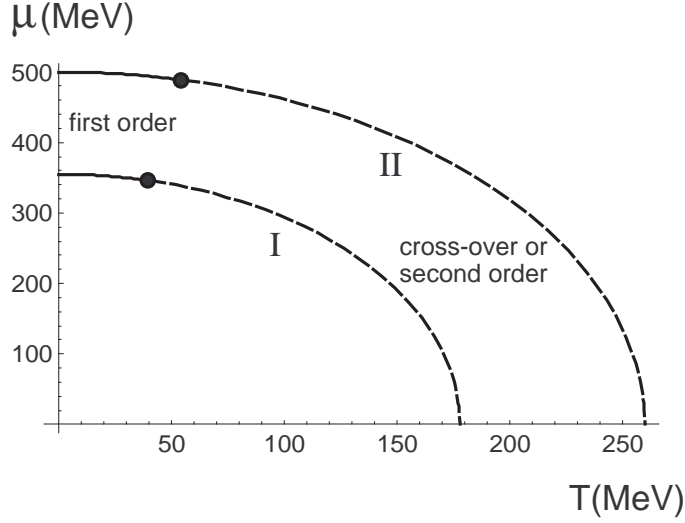}\end{center}
\caption{The continuous lines and the dashed lines correspond to
first-order and to second-order or crossover transitions. The label
I and II refer to different choices of the parameters, see the
text.} \label{fig2}
\end{figure}
The results, in the plane $(T,\mu)$, are shown in Fig. \ref{fig2}
for two different choices of the parameters: I: $m=5.5~MeV$,
$g=5.074\times 10^{-6}~MeV^{-2}$, $\Lambda= 631~MeV$; II:
$m=5.0~MeV$, $g=2.337\times 10^{-6}~MeV^{-2}$, $\Lambda= 925~MeV$.
The corresponding values of the condensates are in the first case
$\langle\bar\psi\psi\rangle= (-247)^3~MeV^3$ and in the second one
$\langle\bar\psi\psi\rangle= (-359)^3~MeV^3$. In the figure we see
the occurrence of a critical end point where the first order
transition line ends.

Refs. \cite{Barducci:1989wi,Barducci:1989eu}  developed an
approximation scheme to QCD (today known as ladder QCD) by using the
Cornwall, Jackiw and Tomboulis (CJT) effective action
\cite{Cornwall:1974vz}. The calculation was done at two loops and it
is equivalent to sum up the ladder diagrams with gluon exchanged. A
further approximation was to use an ansatz for the self-energy of
the type \be
\Sigma(p,T,\mu)=\chi(T,\mu)\frac{\Lambda}{\Lambda^2+p^2}\ee in order
to provide an asymptotic behavior consistent with the operator
product expansion. $\Lambda$ is a mass scale parameter and
$\chi(T,\mu)$ is determined by minimization of the CJT effective
potential. The fermionic condensate is related to $\chi(T,\mu)$ by
the relation \be
\langle\bar\psi\psi\rangle_{T,\mu}=3\frac{\Lambda^3}{g^2(T,\mu)}\chi(T,\mu)\ee
\begin{figure}[h]\begin{center}
\includegraphics[width=.8\textwidth]{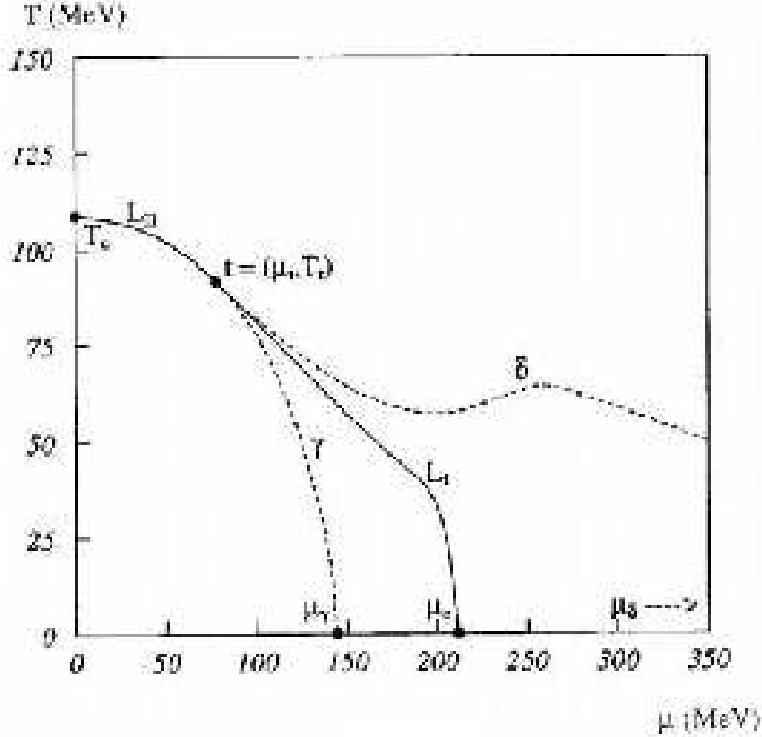}\end{center}
\caption{The continuous lines $L_I$ and $L_{II}$  correspond to the
first and second order transitions respectively. The dot is the
tricritical point, designed as $(\mu_t, T_t)$. The dashed lines
$\gamma$ and $\delta$ (spinodal lines) are explained in the text.}
\label{fig3}
\end{figure}
At $T=\mu=0$, $g(T,\mu)$ is the QCD gauge coupling. The dependence
of $g$ on $T$ and $\mu$ was chosen in a way consistent with
asymptotic freedom \cite{Barducci:1989wi}. At $T=\mu=0$ the
parameters were chosen to be $\Lambda=282~MeV$, $\alpha_s=0.902$. In
this way $\langle\bar\psi\psi\rangle$, renormalized at the scale
$\Lambda$, turns out to be $(-197~MeV)^3$. The results of this
analysis are shown in Fig. \ref{fig3}. The lines denoted by $L_{I}$
and $L_{II}$ correspond to first-order and second-order transitions
(in these papers the massless quark case was considered). These two
lines are separated by the critical end point (a tricritical point
in this case). The dashed line $\gamma$ is the continuation of the
points where the second derivative vanishes at the minimum, whereas
the line $\delta$ is the location of the points where the minima of
the potential go from three to  one (see also Fig. \ref{fig4}).
These two lines are called spinodal lines. The regions between
$\gamma$ and $L_{I}$ and $L_I$ and $\delta$ correspond to metastable
states. We see that the qualitative results are very similar to the
ones obtained in \cite{Asakawa:1989bq} for a completely different
model. Since the tricritical point is where second order and first
order transitions meet together, one can perform a Ginzburg-Landau
expansion of the effective potential. This was done in
\cite{Barducci:1990sv}. By performing the expansion up to $6^{th}$
order in the condensate $\chi$ one gets \be
V(\chi,T,\mu)=V(0,T,\mu)+a_2(T,\mu)\chi^2+a_4(T,\mu)\chi^4+
a_6(T,\mu)\chi^6\label{eq3.5}\ee The coefficients $a_i(\chi,T)$ have
been evaluated in \cite{Barducci:1990sv}. From this expression one
easily derives the phase diagram in the plane $(a_2/a_6,a_4/a_6)$.
In fact, it turns out that $a_6$ is a positive definite quantity in
the region around the critical point. The resulting phase diagram is
illustrated in Fig. \ref{fig4} (see ref. \cite{Barducci:1990kv}).
\begin{figure}[h]\begin{center}
\includegraphics[width=.7\textwidth]{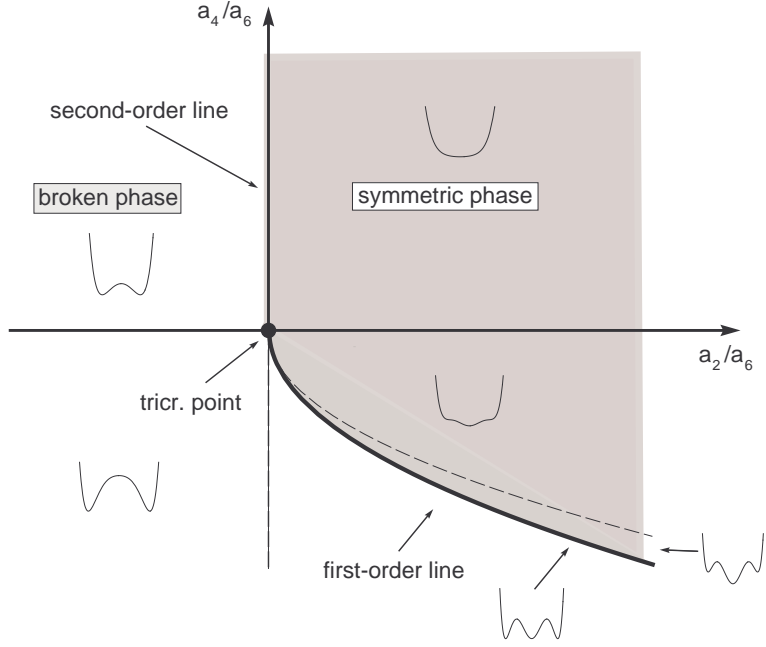}\end{center}
\caption{The phase diagram arising form the Ginzburg-Landau
expansion of the effective potential. The two dashed lines
correspond to the dashed lines $\gamma$ and $\delta$ of the previous
figure. In particular, the line $a_2=0,~a_4\le 0$ corresponds to the
line $\gamma$. The shape of the effective potential in the various
regions is shown in the figure.} \label{fig4}
\end{figure}
The phase diagram of Fig. \ref{fig3} is obtained from this one by
mapping the plane $(a_2/a_6,a_4/a_6)$ into the plane $(\mu, T)$, at
least in the neighborhood of the tricritical point. The second order
line corresponds to $a_2=0$ and $a_4>0$, whereas the tricritical
point is located at $a_2=a_4=0$. We see clearly from Fig. \ref{fig4}
the presence of the metastable regions. By using this approach it is
possible to evaluate the critical exponents around the tricritical
point. In fact, when close to it we can write \be a_i(T,\mu)\approx
a_{iT}\left|\frac{T-T_c}{T_c}\right|+a_{i\mu}\left|
\frac{\mu-\mu_c}{\mu_c}\right|,~~~i=2,4\ee Let us introduce a quark
mass term which, in this variables, is proportional to the field
$\chi$, say \be V_m=-h\chi\ee Then the minimum condition becomes \be
h=2a_2\chi+4a_4\chi^3+6a_6\chi^5\ee and denoting by $\theta$ either
$\mu$ or $T$ one gets \cite{Barducci:1990sv} \bea
\langle\chi\rangle_{m_q=0,\,\theta\to\theta_c}&\approx&
\left|1-\frac\theta{\theta_c}\right|^{1/4}\rightarrow \beta=\frac 1
4\cr \langle\chi\rangle_{m_q\to 0,\,\theta=\theta_c}&\approx&
m_q^{1/5}\rightarrow \delta=\frac 1 5 \cr
\frac{\partial\langle\chi\rangle}{\partial m_q}_{m_q =
0,\,\theta\to\theta_c}&\approx&
\left|1-\frac\theta{\theta_c}\right|^{-1}\rightarrow \gamma=1\eea
where $\alpha$, $\beta$ and $\gamma$ are the usual critical
exponents. Using these relations and the scaling relations for a
three-dimensional system (since the finite temperature cutoff the
time-like modes): \be \alpha=2-3\nu,~~\beta=\frac\nu 2(1+\eta),~~~
\gamma=(2-\eta)\nu,~~~\delta=\frac{5-\eta}{1+\eta}\ee one gets \be
\alpha= \frac 1 2,~~~\nu=\frac 1 2,~~~\eta=0\ee The coefficients
$\alpha$, $\eta$ and $\nu$ define the behavior of the specific heat,
$C(\theta)$, of the correlation length, $\xi(\theta)$, and of the
correlation function at zero momentum, $G(k\to 0)$ \bea
C(\theta)&\approx& \left|1- \frac\theta{\theta_c}\right|^{-\alpha}=
\left|1- \frac\theta{\theta_c}\right|^{-1/2}\cr
\xi(\theta)&\approx&\left|1-\frac\theta{\theta_c}\right|^{-\nu}=
\left|1-\frac\theta{\theta_c}\right|^{-1/2}\eea \be
G_{\alpha\beta}(k\to 0)\approx k^{-2+\eta}=k^{-2}\ee In 1990 when we
got these results  we discovered a paper by Wolff
\cite{Wolff:1985av} showing that the two-dimensional Gross-Neveu
(GN) model has exactly the same phase structure found by us
\cite{Barducci:1989wi}. This was really interesting in view of the
many similarities of the GN model with QCD. This convinced us that,
at least qualitatively, the approximations done in our calculations
did not destroy the main properties of QCD. Also the phase diagram
with a tricritical point occurs in many physical systems. For
instance, in the vapor-liquid transition. However, in this case the
diagram usually plotted is in the variables (density, pressure) or
(volume, pressure). What happens is that the degenerate minima of
the first order line correspond to different densities and the line
$L_I$ splits into two different lines. This is shown in the
(density, pressure) plane in Fig. \ref{fig5} (see, for instance,
ref. \cite{Barducci:1994cb}).
\begin{figure}[h]\begin{center}
\includegraphics[width=.7\textwidth]{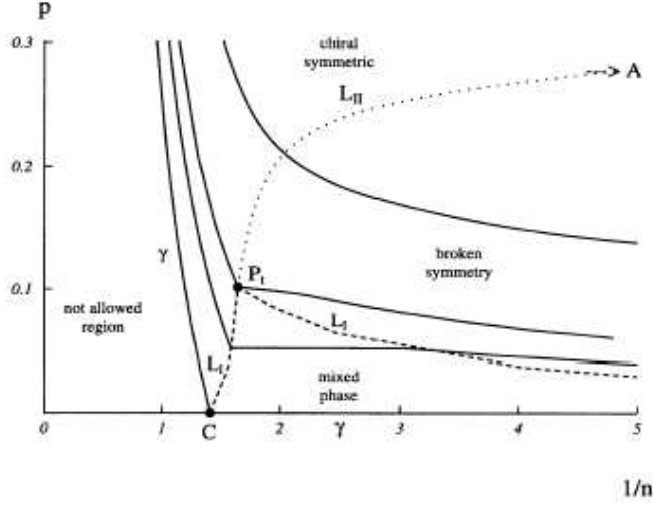}\end{center}
\caption{The phase diagram for ladder QCD (the same as for the
vapor-liquid or for the GN model) in the plane $(1/n,p)$.}
\label{fig5}
\end{figure}

\section{Order  of the transition at zero density vs. the strange quark mass}

In 1984  Pisarski and Wilczek \cite{Pisarski:1983ms} started to
investigate the order of the transition at zero density. In
particular they investigated the dependence of the order on the
number of flavors. They used an effective theory of QCD based on the
introduction of light fields transforming as the chiral condensate:
\be \Phi\approx\bar\psi_L\psi_R\ee When close to the transition,
this is a light field  since the condensate, and therefore the mass
term for $\Phi$, vanishes at $T=T_c$. The transformation properties
of $\Phi$ under the  group $G=U(1)_A\otimes SU(N)_L\otimes SU(N)_R$
are \be \Phi\to e^{i\alpha} U_L\Phi U_R\,,~~~U_L\in
SU(N)_L\,,~U_R\in SU(N)_R\ee It is convenient to parameterize $\Phi$
in the form \be \Phi=\phi U,~~~ U\in SU(N)\ee In this way we
separate the $N^2-1$ Goldstone fields from the condensate $\phi$.
The effective G-invariant lagrangian is \cite{Pisarski:1983ms} \be
{\cal L}=\frac 1 2 {\rm
tr}\,(\partial_\mu\Phi^\dagger\partial^\mu\Phi)- \frac 1 2 m^2{\rm
tr}\,(\Phi^\dagger\Phi)-g_1\left(tr\,(\Phi^\dagger\Phi)\right)^2-
g_2{\rm tr}\,(\Phi^\dagger\Phi)^2\label{eq4.4}\ee To this
G-invariant part a piece breaking $U(1)_A\to Z_A(N)$ is added \be
{\cal L}^\prime=c\left({\rm det}\,\Phi+{\rm
det}\,\Phi^\dagger\right)\label{eq4.5}\ee At zero temperature the
symmetry breaking to $SU(N)_{L+R}$ is enforced through the non
vanishing expectation value \be \langle\Phi\rangle=\Phi_0 \cdot 1\ee
In \cite{Pisarski:1983ms}  the $\beta$-function of this effective
theory has been studied and it was found that for $N\ge 3$ the
transition to the symmetric phase is first-order. The proof goes
through the use of the $\epsilon$-expansion in $4-\epsilon$
dimensions and the analytic continuation to $\epsilon =1$. In this
way one takes into account that, due to the thermal cutoff of the
time-like modes, the theory is effectively three dimensional. About
this point notice that at $d=3$ the dimensions of the  scalar fields
are $[\Phi]=1/2$. As a consequence the sixth order term $\phi^6$ is
marginal and it should be included into the effective expansion when
the coefficients of $\Phi^2$ and $\Phi^4$ are small, that is around
the tricritical point. At zero density and close to the critical
point $T_c$ the effective potential for the condensate $\phi$ can be
taken of of the form (see eqs. \ref{eq4.4} and \ref{eq4.5}) \be
\frac{M^2} 2\phi^2+\frac\lambda 4\phi^4\label{eq4.7}\ee with the
coefficients $M^2$ and $\lambda$ depending on the parameter of the
effective lagrangian $m^2$, $g_1$, $g_2$, $c$ and $N$. The critical
temperature can then determined as a function of the parameters by
the equation \be M(T_c)=0\ee This result was also confirmed by
lattice calculations, as shown in Table \ref{table1}.

\begin{figure}[h]\begin{center}
\includegraphics[width=.5\textwidth]{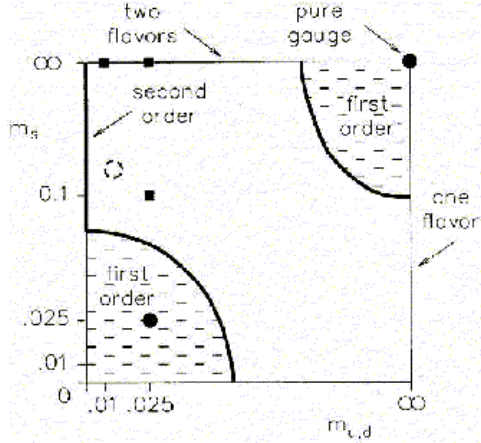}\end{center}
\caption{Solid circles is where the first order transition is seen,
whereas crossover transitions correspond to the solid squares. The
dashed circle indicates  the physical point.} \label{fig6}
\end{figure}

\begin{table}[h]\begin{center}
\begin{tabular}{|c|c|c|c|c|c|c|}
  \hline
  Date  & Authors &  $N=2$ & $N=3$ & $N=4$ & $N=6$ & Lattice size\\
  \hline
  1987 & Gottlieb et al.  \cite{Gottlieb:1987en} & crossover &  & $1^{st}$ &  & $(8,10)^3\times 4$ \\
  1990 & Gottlieb et al. \cite{Gottlieb:1989mh}& crossover &  &  &  & $12^3\times 8$ \\
  1990 & Fukugita et al. \cite{Fukugita:1990vu} & crossover &  & $1^{st}$ &  & $12^3\times 4$\\
  1990 & Kogut et al. \cite{Kogut:1990ii} & crossover & $1^{st}$? & $1^{st}$ &  & $12^3\times 4$ \\
  1990 & Brown et al. \cite{Brown:1990ev} & crossover & $1^{st}$ &  &  & $16^3\times 4$ \\
  1992 & Bernard et al. \cite{Bernard:1991np} & crossover &  &  &  & $12^3\times 6$ \\
  1994 & Zhu \cite{Zhu:1994sj} & crossover   &  &  &  & $(16,32)^3\times 8$ \\
  1995 & Iwasaki et al. \cite{Iwasaki:1995ij} & crossover & $1^{st}$ &  & $1^{st}$ & $12^3\time6(6,18),
  18^2\times 24 \times (6,18)$\\
  \hline
\end{tabular}\caption{A compilation of the results obtained
in lattice calculations of the order of the chiral transition at
$\mu=0$ and $T\not=0$.}\label{table1}
\end{center}
\end{table}
 Following ref. \cite{Brown:1990ev} we may draw the phase diagram in
the space of the quark masses as in Fig. \ref{fig6}.
By looking at this diagram an interesting question arises: how do we
go from a second order to a first order phase transition by varying
the strange quark mass? An  answer to this question was given in
refs. \cite{Wilczek:1992sf,Rajagopal:1992qz}. The argument is the
following: adding a massive quark does not change the effective
action since the light fields are unchanged. Therefore, when close
to the critical point, the effective potential is still of the form
given in eq. \ref{eq4.7}. However the massive quark renormalizes the
couplings. The resulting effect is that a variation of $M^2$ will
change  the critical temperature. However $\lambda$ could change in
such a way to go through zero and become negative. If this is the
case we know that we have to add a $\phi^6$ term in the potential
(remember that the effective theory is three dimensional and that
such an operator is marginal). In this way we may go smoothly from a
second order to a first order transition.

\section{Universality at non zero density}
\begin{figure}[h]\begin{center}
\includegraphics[width=.6\textwidth]{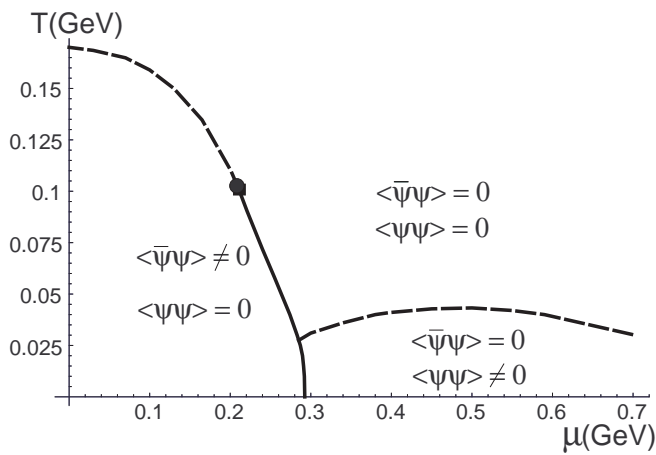}\end{center}
\caption{Dashed lines represent second order phase transitions,
whereas the solid line corresponds to a first order one. The solid
circle indicates the tricritical point.} \label{fig7}
\end{figure}
At the end of the 90's there was a big revival of the studies of the
phase diagram of QCD prompted by the analysis made at very large
density and zero temperature \cite{cs}, showing the formation of a
diquark condensate $\langle\psi\psi\rangle$ and the corresponding
breaking of the color symmetry. After that Berges and Rajagopal
\cite{Berges:1998rc} studied the coexistence of the chiral and of
the diquark condensates in a NJL model. The phase diagram found by
these authors is shown in Fig. \ref{fig7}, and it shows the presence
of the tricritical point. The authors justified the presence of the
tricritical point by using an argument very close to the one used in
refs. \cite{Wilczek:1992sf,Rajagopal:1992qz} in the case of  the
strange quark that we have discussed in the previous section. The
idea is that at zero quark mass the theory belongs to the O(4)
universality class (Ising $Z_2$ for $m_q \not= 0$) and this is not
changed at finite density as shown in ref. \cite{Hsu:1998eu}.
However the renormalization of the coefficients in the effective
action due to the presence of the chemical potential might change
the coefficient of the quartic term forcing the introduction of a
sixth order term in the potential. Again, the $6^{\,th}$ order term
gives rise to a tricritical point in the phase diagram. Also,
recalling that the $\phi^6$ operator is marginal, one expects, at
most, logarithmic corrections  to the critical exponents  evaluated
before.

\begin{figure}[h]\begin{center}
\includegraphics[width=.6\textwidth]{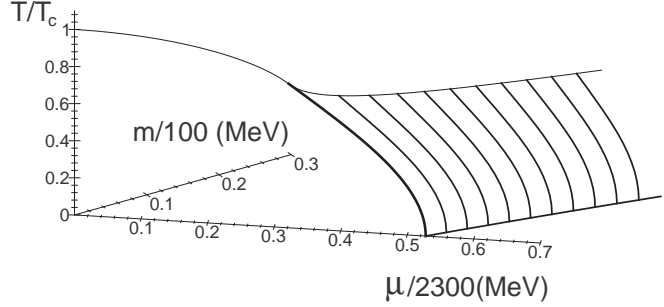}\end{center}
\caption{We show the first order lines as curves at constant quark
mass, $m$. At $m=0$ the second order line is shown. For $m\not=0$
the first order line ends into a critical end point. } \label{fig8}
\end{figure}

After the previous paper many authors reconsidered the problem of
QCD at finite density and temperature using many different
approaches. We will give here a brief list of papers delaing with
the problem.  In 1998 Halasz et al. \cite{Halasz:1998qr} considered
a random matrix model, for the two-flavor case, in the space
$(\mu,T,m)$ finding results consistent the universality arguments.
Their results are shown in Fig. \ref{fig8}.

\begin{figure}[h]\begin{center}
\includegraphics[width=.7\textwidth]{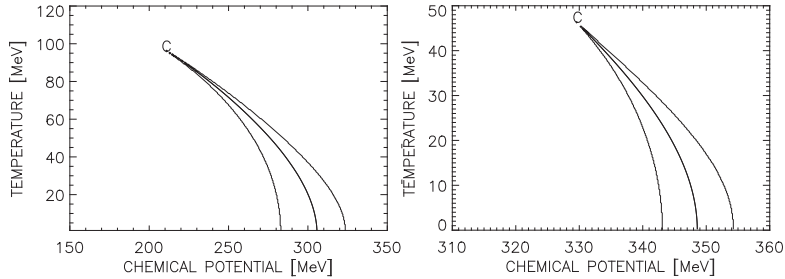}\end{center}
\caption{The two panels illustrate the phase diagram for the
$\sigma$-model (left panel) and for the NJL model (right panel). The
middle curves are the critical lines whereas the outer lines are the
spinodal lines.} \label{fig9}
\end{figure}

In ref. \cite{Scavenius:2000qd} the chiral phase transition has been
examined both in a linear $\sigma$-model and in a NJL model. The two
cases are illustrated in Fig. \ref{fig9}. Once again the results
agree very well with the universality arguments.

\begin{figure}[h]\begin{center}
\includegraphics[width=.55\textwidth]{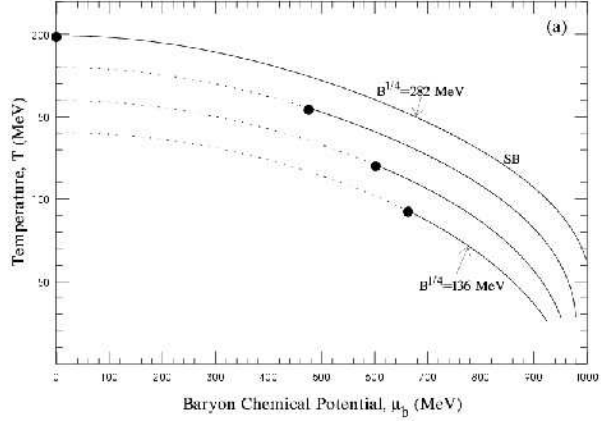}\end{center}
\caption{The $(\mu,T)$ phase diagram from the statistical boostrap
model. $B$ is the value of the bag constant.} \label{fig10}
\end{figure}

Another different approach was considered in ref.
\cite{Antoniou:2002xq}. The calculation was done within the context
of the statistical boostrap principle, and again it agrees with the
universality hypothesis, see Fig. \ref{fig10}. It was also found
that the critical chemical potential is non zero for a large range
of values of the bag constant $B$ ($B^{1/4}< 282~MeV$).
\begin{figure}[h]\begin{center}
\includegraphics[width=.5\textwidth]{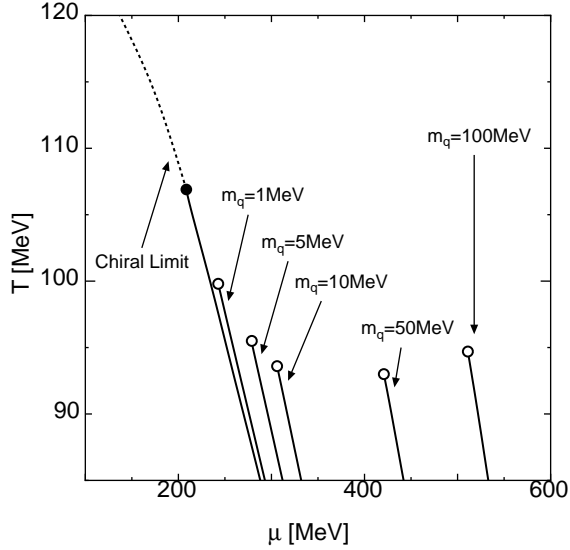}\end{center}
\caption{Quark masses are evaluated at the momentum scale 1 $GeV$.
The solid and dotted lines correspond to first order and second
order phase transitions respectively. The solid and open circles
denotes the tricritical and the critical end points.} \label{fig11}
\end{figure}

As a last example we report a recent calculation made by Hatta and
Ikeda \cite{Hatta:2002sj} using again ladder QCD with the help of
the CJT potential as in refs.
\cite{Barducci:1989wi,Barducci:1989eu}. This calculation takes also
into account  quark masses and it shows the dependence of the
critical end point with this variable, as shown in Fig. \ref{fig11}.

A comparison of the locations of the critical end point evaluated in
different models  can be found in a recent review by Stephanov
\cite{Stephanov:2004wx}. This comparison is particularly interesting
since it shows that, although different models agree qualitatively
well, from a quantitative point of view they are quite different.
For instance, at the critical end point,  the critical value of the
chemical potential varies between roughly 300 $MeV$ up to about 1000
$MeV$, whereas the critical value of the temperature goes between 40
and 170 $MeV$. Clearly in order to have a quantitative improvement
one would need a first principles calculation.

\section{Lattice calculations}

As noticed in the previous section one would really need to have the
possibility of testing on the lattice the phase diagram of QCD.
However the usual sampling method, based on a positive definite
measure in the euclidean path integral, does not work in presence of
a real chemical potential, since the fermionic determinant is then
complex. In fact, let us define euclidean variables through the
following substitutions: \be x_0\to -ix_E^4,~~ x^i\to x^i_E,~~
\gamma_0\to\gamma_E^4, ~~\gamma^i\to -i\gamma_E^i\ee The euclidean
Dirac operator in the presence of a chemical potential is \be
D(\mu)=\gamma^\mu_E D^\mu_E+
\mu\gamma_E^4,~~~D^\mu_E=\partial_E^\mu+iA_E^\mu\ee  At $\mu=0$ the
eigenvalues of $D(\mu)$ are pure imaginary and also, if
$|\lambda\rangle$ is an eigenvector of $D(0)$, then
$\gamma_5|\lambda\rangle$ belongs to the eigenvalue $-\lambda_5$, as
it follows from \be
D(0)^\dagger=-D(0),~~~\gamma_5D(0)\gamma_5=-D(0)\ee Therefore \be
{\rm det}[D(0)]=\prod_{\lambda}(\lambda)(-\lambda)>0\ee At
$\mu\not=0$ this argument does not hold and we lack the positivity
property. However, if one considers the chemical potential
associated to the isospin, since this is related to the conserved
current $\tau_3$, the positivity can be proved by using $\tau_1$ in
conjunction with the hermitian conjugation.

Recently there have been numerous different attempts to improve the
lattice calculations at $\mu\not=0$:
\begin{itemize}
\item Rewighting method (see, for instance \cite{Fodor:2001au,Fodor:2001pe,Fodor:2004nz}).
\item Taylor expansion for small $\mu$ (see, for instance
\cite{Allton:2002zi}, \cite{Allton:2003vx}, \cite{Ejiri:2003dc})
\item Imaginary chemical potential (see, for example
\cite{Lombardo:1999cz}, \cite{D'Elia:2002gd}
\cite{deForcrand:2002ci}, \cite{deForcrand:2003hx},
\cite{Stephanov:2006dn})
\end{itemize}

\subsection{Reweighting}

The reweighting technique (for a review see ref.
\cite{Barbour:1996pn})is based on the following identity for the
partition function \be \int D\,U {\rm det}\,[D(0)]\frac{{\rm
det}\,[D(\mu)]}{{\rm det}\,[D(0)]}e^{-S_g(U)}=\left\langle\frac{{\rm
det}\,[D(\mu)]}{{\rm det}\,[D(0)]}\right\rangle_{\mu=0}\ee
\begin{figure}[h]\begin{center}
\includegraphics[width=.6\textwidth]{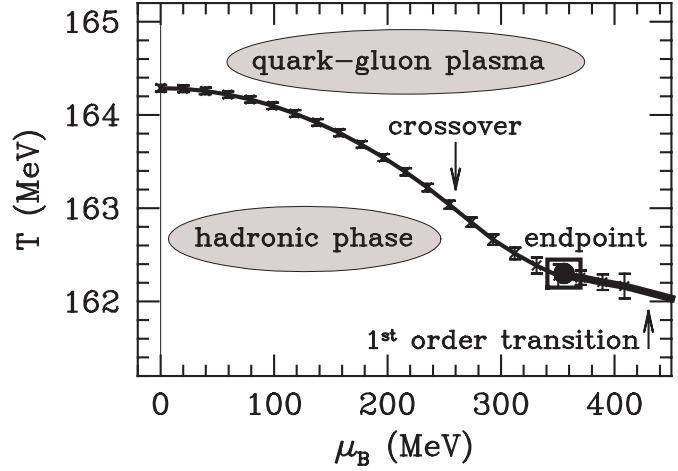}\end{center}
\caption{The most recent determination with the reweighting
procedure is given in ref. \cite{Fodor:2004nz}. The errors are due
to the reweighting procedure and on the error of the scale
determination at $T=0$. The values of the critical temperature and
chemical potential are $T=162\pm 2~MeV$, $\mu=360\pm 40~MeV$.}
\label{fig12}
\end{figure}
Since the integration measure is taken at $\mu=0$ it is positive
definite. However in the numerical calculation problems arise. The
ratio of the two determinants oscillates and there are large
cancellations. Also, since the reweighting corresponds to the ratio
of two partition functions with different actions, it decays
exponentially according to the difference of the free energies,
$\Delta F$. This is proportional to the volume and therefore the
statistics required for a given accuracy increases with the volume.
An improvement of this technique is the so called "multiparameter
reweighting" which is a generalization of the previous method
\cite{Fodor:2001au}. The idea is to reweight also in the lattice
gauge coupling, writing \be Z=\left\langle\frac{e^{-S_g(\beta)}{\rm
det}\,[D(\mu)]}{e^{-S_g(\beta_0)}{\rm
det}\,[D(0)]}\right\rangle_{\mu=0,\beta_0}\ee The second reweighting
parameter can be used to allow the statistical ensemble to fluctuate
between the phases and to avoid that the ensemble goes away from
criticality. One of the most recent calculations using this
technique was done in ref. \cite{Fodor:2004nz} and the result is
shown in Fig. \ref{fig12}

\subsection{Taylor expansion}
\bigskip
\begin{figure}[h]\begin{center}
\includegraphics[width=.6\textwidth]{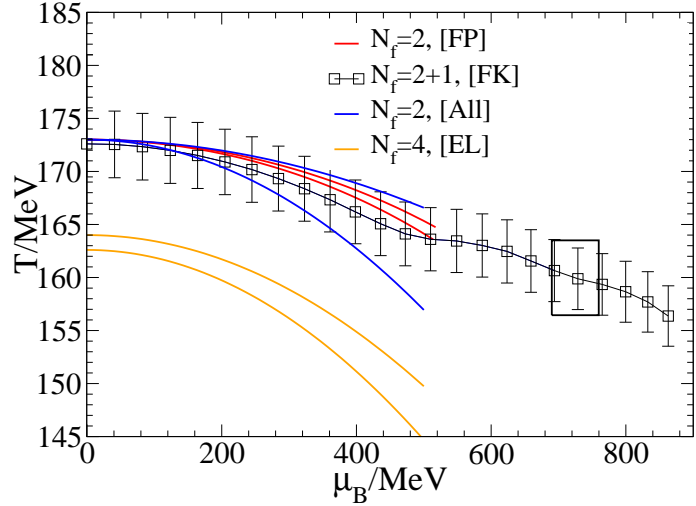}\end{center}
\caption{A comparison of the critical line in different simulations,
FP \cite{deForcrand:2002ci}, FK \cite{Fodor:2001pe}, All
\cite{Allton:2002zi}, EL \cite{D'Elia:2002gd}. The box is the
position of the critical end point evaluated by Fodor and Katz in
\cite{Fodor:2001pe}.} \label{fig13}
\end{figure}
This method makes use of the multiparameter reweighting and at the
same time of a Taylor expansion in the chemical potential. Although
this method is not useful for determining the critical end point, it
is of interest in the heavy ion physics where values of $\mu$ of a
few ten $MeV$ are important. What one does is simply to expand in a
Taylor series of $\mu/T$ the reweighting factor. In particular this
method can be used to evaluate the behavior of the critical line at
small $\mu/T$. This method has been used in ref.
\cite{Allton:2002zi} for the two-flavor case. The results are shown
in Fig. \ref{fig13}.

\subsection{Imaginary chemical potential}

If $\mu$ is pure imaginary the fermion determinant is positive and
numerical simulations can be done easily as for the case $\mu=0$.
Using the fact that the observables are analytic functions of $\mu$
except that on the critical line, one computes expectation values at
imaginary $\mu$ and then one  fits them by a truncated Taylor
expansion \cite{Lombardo:1999cz,D'Elia:2002gd}. Some of these
results are given in Fig. \ref{fig13}.

\section{Isospin chemical potential}

\begin{figure}[h]\begin{center}
\includegraphics[width=1\textwidth]{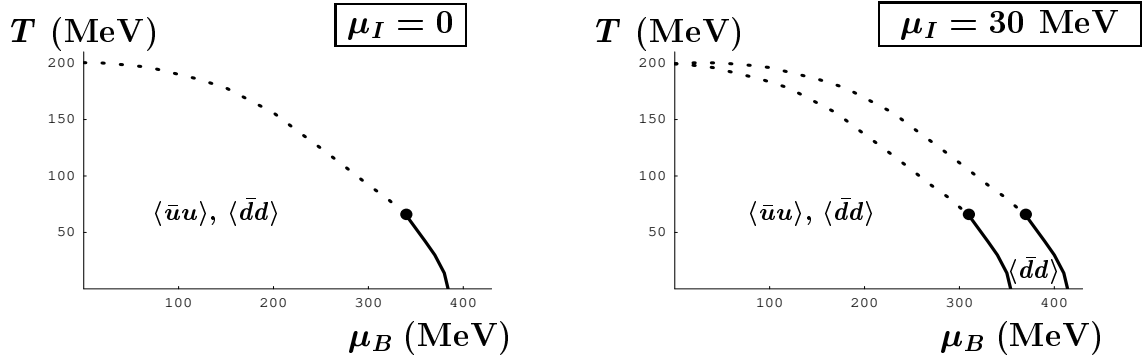}\end{center}
\caption{The phase diagram for the NJL model as studied in ref.
\cite{Toublan:2003tt}. The left panel shows the case $\mu_I=0$,
whereas in the right panel $\mu_I= 30~MeV$.} \label{fig14}
\end{figure}

To end this review we will report also some result obtained  in
presence of an isospin chemical potential, $\mu_I$. The case of
$\mu_I\not=0$ is, in principle, interesting for the heavy ion
physics. It is also interesting since for $\mu=0$ and $\mu_I\not=0$
the fermionic determinant is positive
\cite{Kogut:2002zg,Gupta:2002kp}. The problem ($\mu$ and $\mu_I\not
=0$) has been studied using effective lagrangians
\cite{Son:2000xc,Loewe:2002tw}, random matrices \cite{Klein:2003fy},
NJL model \cite{Toublan:2003tt,Barducci:2004tt} and ladder-QCD
\cite{Barducci:2003un}. The most interesting effect in these studies
appears to be the splitting of the critical line and of the critical
end point. In fact, this effect could bring down the end critical
point to a region more accessible to heavy ion experiments. We show
in Fig. \ref{fig14} the result in the NJL model
\cite{Toublan:2003tt}. The result is qualitatively compatible with
an analogous calculation made in the ladder-QCD model
\cite{Barducci:2003un}. However a too strong  mixing of the up and
down flavors could destroy this interesting result, as shown in ref.
\cite{Frank:2003ve}.

\begin{figure}[h]\begin{center}
\includegraphics[width=.8\textwidth]{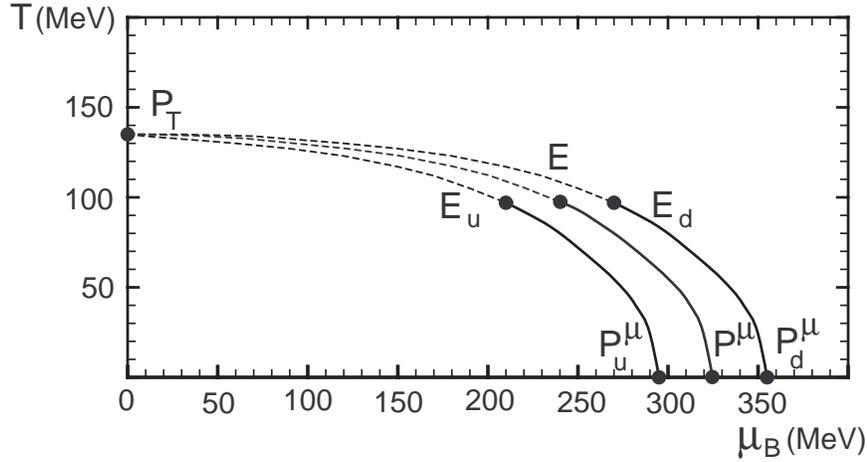}\end{center}
\caption{The phase diagram for the ladder-QCD model as studied in
ref. \cite{Barducci:2003un}.} \label{fig15}
\end{figure}

\section{Conclusions}

As we have shown there has been a lot of activities and results in
our understanding of the phase diagram of QCD. However most of the
progress is still at a very qualitative level. We would like to be
able to locate the critical point with a good accuracy, but for
that, a breakthrough is real necessary. This could come by devising
some clever technique in lattice calculations, or, may be, a new
analytical method.


\begin{thebibliography}{99}

\bibitem{barrois}

B. Barrois, { Nuclear Physics} {\bf B129}, 390 (1977);
S. Frautschi, {\it Proceedings of workshop on hadronic matter at
extreme density}, Erice 1978; D. Bailin and A. Love, { Physics
Report} {\bf 107} (1984) 325 .

\bibitem{cs} M.~Alford, K.~Rajagopal, and F.~Wilczek,
{ Phys.\ Lett.}\  {\bf B422}(1998) 247 [{\tt hep-ph/9711395}];
R.~Rapp, T.~Schafer, E.~V.~Shuryak and M.~Velkovsky,
{Phys.\ Rev.\ Lett.}\  {\bf 81}, 53 (1998) [{\tt hep-ph/9711396}].

\bibitem{Rajagopal:1995bc}
  K.~Rajagopal,
  arXiv:hep-ph/9504310.

\bibitem{Rajagopal:2000wf}
  K.~Rajagopal and F.~Wilczek,
  arXiv:hep-ph/0011333.

\bibitem{Stephanov:2004wx}
  M.~A.~Stephanov,
  Prog.\ Theor.\ Phys.\ Suppl.\  {\bf 153} (2004) 139
  [Int.\ J.\ Mod.\ Phys.\ A {\bf 20} (2005) 4387]
  [arXiv:hep-ph/0402115].

\bibitem{Satz:2005md}
  H.~Satz,
  Int.\ J.\ Mod.\ Phys.\ A {\bf 21} (2006) 672
  [arXiv:hep-lat/0509192].


\bibitem{Schafer:2005ff}
  T.~Schafer,
  arXiv:hep-ph/0509068.


\bibitem{polyakov}
  L.~D.~McLerran and B.~Svetitsky,
  Phys.\ Lett.\ B {\bf 98} (1981) 195;
  L.~D.~McLerran and B.~Svetitsky,
  Phys.\ Rev.\ D {\bf 24} (1981) 450;
J.~Kuti, J.~Polonyi and K.~Szlachanyi,
  Phys.\ Lett.\ B {\bf 98}, 199 (1981).

\bibitem{Karsch:1994hm}
  F.~Karsch and E.~Laermann,
  Phys.\ Rev.\ D {\bf 50} (1994) 6954
  [arXiv:hep-lat/9406008].

\bibitem{Bailin:1984ak}
  D.~Bailin, J.~Cleymans and M.~D.~Scadron,
  Phys.\ Rev.\ D {\bf 31} (1985) 164.

\bibitem{Kocic:1985uq}
  A.~Kocic,
  Phys.\ Rev.\ D {\bf 33} (1986) 1785.

\bibitem{Galina:1987bj}
  V.~F.~Galina and K.~S.~Viswanathan,
  Phys.\ Rev.\ D {\bf 38} (1988) 2000.

\bibitem{Damgaard:1985bn}
  P.~H.~Damgaard, D.~Hochberg and N.~Kawamoto,
  Phys.\ Lett.\ B {\bf 158} (1985) 239.

\bibitem{Ilgenfritz:1984ff}
  E.~M.~Ilgenfritz and J.~Kripfganz,
  Z.\ Phys.\ C {\bf 29} (1985) 79.

\bibitem{Asakawa:1989bq}
  M.~Asakawa and K.~Yazaki,
  Nucl.\ Phys.\ A {\bf 504} (1989) 668.

\bibitem{Barducci:1989wi}
  A.~Barducci, R.~Casalbuoni, S.~De Curtis, R.~Gatto and G.~Pettini,
  Phys.\ Lett.\ B {\bf 231} (1989) 463.

\bibitem{Barducci:1989eu}
  A.~Barducci, R.~Casalbuoni, S.~De Curtis, R.~Gatto and G.~Pettini,
  Phys.\ Rev.\ D {\bf 41} (1990) 1610.

\bibitem{Cornwall:1974vz}
  J.~M.~Cornwall, R.~Jackiw and E.~Tomboulis,
  Phys.\ Rev.\ D {\bf 10}, 2428 (1974).

\bibitem{Barducci:1990sv}
  A.~Barducci, R.~Casalbuoni, S.~De Curtis, R.~Gatto and G.~Pettini,
  Phys.\ Rev.\ D {\bf 42} (1990) 1757.

\bibitem{Barducci:1990kv}
  A.~Barducci, R.~Casalbuoni, S.~De Curtis, R.~Gatto and G.~Pettini,
UGVA-DPT-1990-10-697,
Proceedings of the {\it Large hadron collider workshop}, Aachen,
Germany, Oct 4-9, 1990, edited by G. Jarlskog and D. Rein. Geneva,
Switzerland, CERN, 1990. 3 volumes. (CERN-90-10)

\bibitem{Wolff:1985av}
  U.~Wolff,
  Phys.\ Lett.\ B {\bf 157} (1985) 303.

\bibitem{Barducci:1994cb}
  A.~Barducci, R.~Casalbuoni, M.~Modugno, G.~Pettini and R.~Gatto,
  Phys.\ Rev.\ D {\bf 51} (1995) 3042
  [arXiv:hep-th/9406117].

\bibitem{Pisarski:1983ms}
  R.~D.~Pisarski and F.~Wilczek,
  Phys.\ Rev.\ D {\bf 29} (1984) 338.

\bibitem{Gottlieb:1987en}
  S.~A.~Gottlieb, W.~Liu, D.~Toussaint, R.~L.~Renken and R.~L.~Sugar,
  Nucl.\ Phys.\ Proc.\ Suppl.\  {\bf 4} (1988) 155.

\bibitem{Gottlieb:1989mh}
  S.~A.~Gottlieb, W.~Liu, R.~L.~Renken, R.~L.~Sugar and D.~Toussaint,
  Phys.\ Rev.\ D {\bf 41} (1990) 622.

\bibitem{Fukugita:1990vu}
  M.~Fukugita, H.~Mino, M.~Okawa and A.~Ukawa,
  Phys.\ Rev.\ Lett.\  {\bf 65} (1990) 816.

\bibitem{Kogut:1990ii}
  J.~B.~Kogut and D.~K.~Sinclair,
  Nucl.\ Phys.\ B {\bf 344} (1990) 238.

\bibitem{Brown:1990ev}
  F.~R.~Brown {\it et al.},
  Phys.\ Rev.\ Lett.\  {\bf 65} (1990) 2491.

\bibitem{Bernard:1991np}
  C.~W.~Bernard {\it et al.},
  Phys.\ Rev.\ D {\bf 45} (1992) 3854.

\bibitem{Zhu:1994sj}
  D.~c.~Zhu,
   {\it Numerical Study Of Two-Flavor QCD Phase Structures At N(T) = 8 On 16**3 And
  32**3 Volumes}, PhD thesis,
UMI-95-16217

\bibitem{Iwasaki:1995ij}
  Y.~Iwasaki, K.~Kanaya, S.~Sakai and T.~Yoshie,
  Z.\ Phys.\ C {\bf 71} (1996) 337
  [arXiv:hep-lat/9504019].

\bibitem{Wilczek:1992sf}
  F.~Wilczek,
  Int.\ J.\ Mod.\ Phys.\ A {\bf 7} (1992) 3911
  [Erratum-ibid.\ A {\bf 7} (1992) 6951].

\bibitem{Rajagopal:1992qz}
  K.~Rajagopal and F.~Wilczek,
  Nucl.\ Phys.\ B {\bf 399} (1993) 395
  [arXiv:hep-ph/9210253].

\bibitem{Berges:1998rc}
  J.~Berges and K.~Rajagopal,
  Nucl.\ Phys.\ B {\bf 538} (1999) 215
  [arXiv:hep-ph/9804233].

\bibitem{Hsu:1998eu}
  S.~D.~H.~Hsu and M.~Schwetz,
  Phys.\ Lett.\ B {\bf 432} (1998) 203
  [arXiv:hep-ph/9803386].

\bibitem{Halasz:1998qr}
  M.~A.~Halasz, A.~D.~Jackson, R.~E.~Shrock, M.~A.~Stephanov and J.~J.~M.~Verbaarschot,
  Phys.\ Rev.\ D {\bf 58} (1998) 096007
  [arXiv:hep-ph/9804290].

\bibitem{Scavenius:2000qd}
  O.~Scavenius, A.~Mocsy, I.~N.~Mishustin and D.~H.~Rischke,
  Phys.\ Rev.\ C {\bf 64} (2001) 045202
  [arXiv:nucl-th/0007030].

\bibitem{Antoniou:2002xq}
  N.~G.~Antoniou and A.~S.~Kapoyannis,
  Phys.\ Lett.\ B {\bf 563} (2003) 165
  [arXiv:hep-ph/0211392].

\bibitem{Hatta:2002sj}
  Y.~Hatta and T.~Ikeda,
  Phys.\ Rev.\ D {\bf 67} (2003) 014028
  [arXiv:hep-ph/0210284].

\bibitem{Fodor:2001au}
  Z.~Fodor and S.~D.~Katz,
  Phys.\ Lett.\ B {\bf 534} (2002) 87
  [arXiv:hep-lat/0104001].

\bibitem{Fodor:2001pe}
  Z.~Fodor and S.~D.~Katz,
  JHEP {\bf 0203} (2002) 014
  [arXiv:hep-lat/0106002].

\bibitem{Fodor:2004nz}
  Z.~Fodor and S.~D.~Katz,
  JHEP {\bf 0404} (2004) 050
  [arXiv:hep-lat/0402006].

\bibitem{Allton:2002zi}
  C.~R.~Allton {\it et al.},
  Phys.\ Rev.\ D {\bf 66} (2002) 074507
  [arXiv:hep-lat/0204010].

\bibitem{Allton:2003vx}
  C.~R.~Allton, S.~Ejiri, S.~J.~Hands, O.~Kaczmarek, F.~Karsch, E.~Laermann and C.~Schmidt,
  Phys.\ Rev.\ D {\bf 68} (2003) 014507
  [arXiv:hep-lat/0305007].

\bibitem{Ejiri:2003dc}
  S.~Ejiri, C.~R.~Allton, S.~J.~Hands, O.~Kaczmarek, F.~Karsch, E.~Laermann and C.~Schmidt,
  Prog.\ Theor.\ Phys.\ Suppl.\  {\bf 153} (2004) 118
  [arXiv:hep-lat/0312006].

\bibitem{Lombardo:1999cz}
  M.~P.~Lombardo,
  Nucl.\ Phys.\ Proc.\ Suppl.\  {\bf 83} (2000) 375
  [arXiv:hep-lat/9908006].

\bibitem{D'Elia:2002gd}
  M.~D'Elia and M.~P.~Lombardo,
  Phys.\ Rev.\ D {\bf 67} (2003) 014505
  [arXiv:hep-lat/0209146].

\bibitem{deForcrand:2002ci}
  P.~de Forcrand and O.~Philipsen,
  Nucl.\ Phys.\ B {\bf 642} (2002) 290
  [arXiv:hep-lat/0205016].

\bibitem{deForcrand:2003hx}
  P.~de Forcrand and O.~Philipsen,
  Nucl.\ Phys.\ B {\bf 673} (2003) 170
  [arXiv:hep-lat/0307020].

\bibitem{Stephanov:2006dn}
  M.~A.~Stephanov,
  Phys.\ Rev.\ D {\bf 73} (2006) 094508
  [arXiv:hep-lat/0603014].

\bibitem{Barbour:1996pn}
  I.~M.~Barbour, S.~E.~Morrison and J.~B.~Kogut  [UKQCD Collaboration],
  Nucl.\ Phys.\ Proc.\ Suppl.\  {\bf 63} (1998) 436
  [arXiv:hep-lat/9612012].

\bibitem{Kogut:2002zg}
  J.~B.~Kogut and D.~K.~Sinclair,
  Phys.\ Rev.\ D {\bf 66} (2002) 034505
  [arXiv:hep-lat/0202028].

\bibitem{Gupta:2002kp}
  S.~Gupta,
  arXiv:hep-lat/0202005.

\bibitem{Son:2000xc}
  D.~T.~Son and M.~A.~Stephanov,
  Phys.\ Rev.\ Lett.\  {\bf 86}, 592 (2001)
  [arXiv:hep-ph/0005225].

\bibitem{Loewe:2002tw}
  M.~Loewe and C.~Villavicencio,
  Phys.\ Rev.\ D {\bf 67} (2003) 074034
  [arXiv:hep-ph/0212275].

\bibitem{Klein:2003fy}
  B.~Klein, D.~Toublan and J.~J.~M.~Verbaarschot,
  Phys.\ Rev.\ D {\bf 68} (2003) 014009
  [arXiv:hep-ph/0301143].

\bibitem{Toublan:2003tt}
  D.~Toublan and J.~B.~Kogut,
  Phys.\ Lett.\ B {\bf 564} (2003) 212
  [arXiv:hep-ph/0301183].

\bibitem{Barducci:2004tt}
  A.~Barducci, R.~Casalbuoni, G.~Pettini and L.~Ravagli,
  Phys.\ Rev.\ D {\bf 69} (2004) 096004
  [arXiv:hep-ph/0402104].

\bibitem{Barducci:2003un}
  A.~Barducci, G.~Pettini, L.~Ravagli and R.~Casalbuoni,
  Phys.\ Lett.\ B {\bf 564} (2003) 217
  [arXiv:hep-ph/0304019].

\bibitem{Frank:2003ve}
  M.~Frank, M.~Buballa and M.~Oertel,
  Phys.\ Lett.\ B {\bf 562} (2003) 221
  [arXiv:hep-ph/0303109].

\end{thebibliography}
\end{document}